\documentclass{aa}
\usepackage{psfig}
\begin{document}
    \title{Cosmic Background dipole measurements with Planck-High 
    Frequency Instrument}
    
    \author{M. Piat \inst{1} \and G. Lagache 
    \inst{1} \and J.P. Bernard \inst{1} \and M. Giard \inst{2} \and J.L. Puget \inst{1} }
    \offprints{M. Piat, \email{michel.piat@ias.u-psud.fr}}

    \institute{
    Institut d'Astrophysique Spatiale (IAS), Universit\'e 
    Paris Sud, B\^{a}t. 121, 91405 Orsay Cedex, FRANCE
    \and
    Centre d'Etude Spatiale des Rayonnements (CESR), 
    9 avenue du Colonel Roche 
    BP 4346, 31028 Toulouse Cedex4, FRANCE}
    \date{Received October 19, 2001; accepted ??? ??, 200?}
\abstract{  
This paper discusses the Cosmic Background (CB) dipoles observations
in the framework of the Planck mission. Dipoles observations can be
used in three ways: (i) It gives a measurement of the peculiar 
velocity of our Galaxy which is an important observation in large 
scale structures formation model. (ii) Measuring
the dipole can give unprecedent information on the monopole
(that can be in some cases hard to obtain due
to large foreground contaminations). (iii) The dipole
can be an ideal absolute calibrator, easily detectable in
cosmological experiments. \\
Following the last two objectives, the main goal of the work presented here
is twofold. First, we study the accuracy of the Planck-HFI calibration
using the Cosmic Microwave Background (CMB) dipole measured by 
COBE as well as the Earth orbital motion dipole.
We show that we can reach for HFI,
a relative calibration between rings of about 1\% and an absolute calibration
better than 0.4\% for the CMB channels
(in the end, the absolute calibration will
be limited by the uncertainties on the CMB temperature).
We also show that Planck will be able to measure the CMB dipole
direction at better than 1.7 arcmin and improve on the amplitude.
Second, we investigate the detection of the Cosmic Far-Infrared
Background (FIRB) dipole. Measuring this dipole could give a new and
independent determination of the FIRB for which a direct determination
is quite difficult due to Galactic dust emission contamination. We show
that
such a detection would require a Galactic dust emission removal
at better than 1\%, which will be very hard to achieve.
}
\maketitle
%
\section{Introduction: the Cosmic Background and its dipole signal}
\begin{figure}
   \hspace{0cm}\psfig{figure=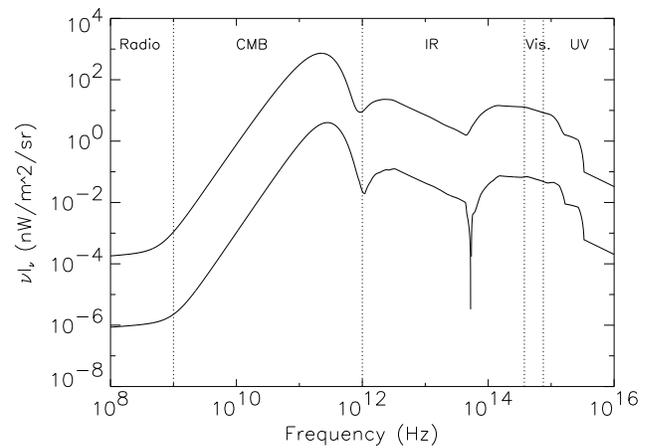,width=8.8cm,angle=90}
   \caption[]{Spectrum of the CB (top) and its dipole 
   amplitude (bottom) from radio to ultra-violet.}
    \label{FDE}
\end{figure}

\begin{figure}
   \hspace{0cm}\psfig{figure=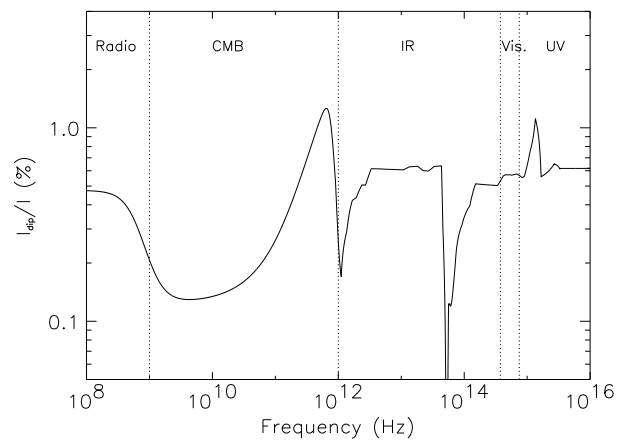,width=8.8cm,angle=90}
   \caption[]{Amplitude of the CB dipole relative to the CB intensity 
   expressed in \% from radio to ultra-violet.}
    \label{IdipsurI}
\end{figure}

The Cosmic Background (CB) is the extragalactic part of the diffuse
electromagnetic emission at all wavelengths.  If the universe obeys
the cosmological principle and is homogeneous and isotropic, this
background is expected to be nearly isotropic in the rest frame where 
the matter in a large volume around the observer has no bulk velocity 
other than the Hubble expansion.  The measurement of the
intensity and Spectral Energy Distribution (SED) of this isotropic
background is a difficult observational challenge.  It requires a
separation of this extragalactic component from all other diffuse
foregrounds (interplanetary or interstellar emission).  Although the
CB was detected quite early in the radio 
(radio-galaxies), Xrays and
gamma rays as it dominates over the foregrounds, these components
account for only 0.027\% of the electromagnetic content of the
present universe.  The CB is dominated by the Cosmic
Microwave Background (CMB) which accounts for 93\% of electromagnetic
content of the present universe and is a truly diffuse component
coming from the pregalactic era of the universe.  The second component
in energy content is the radiation from galaxies integrated over all 
redshifts in the
ultraviolet-optical-near-infrared (stellar emission) and in the
thermal and far infrared (re-emission of stellar radiation absorbed by
interstellar dust). 
This component has been detected only recently, first in the far
infrared in the COBE-FIRAS and COBE-DIRBE data (Puget et al. 1996; Fixsen 
et al. 1998; Hauser et al. 1998; Lagache et al. 2000) then in the optical with HST
(Bernstein et al. (in preparation, quoted by Madau \& Pozzetti 2000); Pozzetti et al. 
1998) and near infrared (Gorjian at al. 2000; Dwek \& Arendt 1999; Cambr\'esy et al. 
2000; Wright 2000).  These measurements have been complemented 
by very high-energy gamma rays absorption by this CB 
(see for example Renault et al. 2001) but are still affected
by rather large uncertainties especially in the optical and mid
infrared.  Figure \ref{FDE} shows the spectrum of the CB from
radio to gamma rays (data from Gispert et al. 2000 for the part 
[30nm; 1mm] and from Halpern \& Scott 1999 for the radio part).\\
The CB should display a dipole anisotropy associated
with the peculiar motion of the observer with respect to the local
cosmological standard of rest.  This peculiar motion has been first
detected for the CMB by Kogut et al. (1993) .  It is composed of several
terms: the peculiar motion of the Sun (sum of the Galaxy peculiar velocity and
the velocity of the Sun with the rotation of our Galaxy), the Earth
velocity in its orbit around the Sun and the specific motion of the
observer with respect to the Earth (orbital motion of the satellite
for example).  The main term is the first one with about 369 km/s in the
direction $(l,b)\simeq (264.3\degr ,\,48.0\degr )$ (Lineweaver et al. \cite{lineweaver}). 
The second one, dominated by the Earth orbital motion, is a 
sinusoidal term with amplitude of about 37 km/s . It is thus about 10\% of the first one but
has a very specific character it is perfectly known in amplitude and
direction and changes periodically with time during the year.
The COBE-DMR experiment has measured these terms accurately. 
Fixsen et al. (1996) has shown that the CMB spectrum
absolutely measured with the COBE-FIRAS experiment with high accuracy
could also be measured using its dipole anisotropy. The two spectra 
obtained were compatible within error bars.  The
peculiar motion of the galaxy has also been measured with respect to
distant galaxies as an anisotropy in the Hubble constant (Lauer \& 
Postman \cite{lauer}). Although the accuracy is not very good, 
the value obtained is compatible with the one measured by COBE 
(Scaramella et al. 1991, Kogut et al. 1993). 
These measurements confirm that the rest frame defined as the one in
which the CMB is isotropic coincides with the rest frame defined by
the galaxies on large scales. \\
Motion with velocity $\vec{\beta}=\vec{v}/c$ through an 
isotropic radiation field of intensity $I(\nu)$ yields an observed 
intensity given by:
\begin{equation}
    I_{\mathrm{obs}}(\nu,\theta)=\frac{I[\nu (1-\beta \cos \theta)]}{(1-\beta \cos 
    \theta)^{3}}
    \label{eq:1}
\end{equation}
where $\nu$ is the frequency and $\theta$ is 
the angle between $\vec{\beta}$ and the direction of observation, 
these two parameters beeing measured in the observer's frame. 
The spectral intensity of the dipole 
amplitude is therefore given by:
\begin{eqnarray}
    I_{\mathrm{dip}}(\nu) & = & I_{\mathrm{obs}}(\nu,\theta=0\degr)-
    I_{\mathrm{obs}}(\nu,\theta=90\degr) \nonumber 
    \\
    & \simeq & \beta \cos \theta \left( 3I(\nu)-\nu 
    \frac{\mathrm{d}I}{\mathrm{d}\nu}(\nu) 
    \right)\;\mathrm{for}\;\beta\ll1
    \label{eq:2}
\end{eqnarray}
The spectral intensity of the CB dipole amplitude is shown in figure 
\ref{FDE} and its relative level in figure \ref{IdipsurI}.\\
The dipole anisotropy of the CB has a two-fold
interest for future CB projects:  
\begin{itemize}
    \item At frequencies where the separation of the CB 
    from foregrounds is difficult
    (from submillimeter to ultraviolet), an accurate measurement of the
    dipole anisotropy will give the SED of the
    CB independently of the interstellar and interplanetary
    foregrounds which do not show a dipole anisotropy. 
    \item A very
    accurate measurement of the CMB dipole anisotropy over a one year
    period should allow to use the dipole anisotropy associated with the
    orbital motion of the Earth around the Sun as the best photometric
    calibrator for extended sources in the microwave region of the 
    spectrum. It should therefore improve the determination of the
    peculiar motion of the solar system.  This is possible due to the
    Planckian nature of the CMB and the accuracy of its temperature
    measured with the FIRAS spectrometer.  The CMB provides today an
    extended photometric standard with an accuracy of less than 
    $2\;10^{-3}$ in thermodynamic temperature (Fixsen et al. 1996).
\end{itemize}
This paper explores these two questions in the context of the Planck
mission and specifically its High Frequency Instrument (Planck-HFI, 
Lamarre \cite{lamarre0}). 
It is divided in 2 parts. The first one deals with the 
calibration of Planck-HFI using the CMB dipole. In  section 
2, we present the calibration philosophy of Planck, while section 
3 concerns more specifically the calibration accuracy. The second 
part, in section 4, is a study of the detectability of the Far InfraRed Background 
(FIRB) dipole. 
\section{Planck-HFI calibration philosophy}
Different sources will be used to monitor inflight the response or to 
get the photometric calibration of the Planck-HFI channels: 
\begin{itemize}
    \item Extended sources: the CMB dipole, the Earth orbital motion 
    dipole, the Galactic disc and eventually maps of Herschel-SPIRE. 
    \item Point sources: planets, infrared galaxies and asteroids. 
\end{itemize}
A general principle 
throughout the calibration for all channels is that relative calibrations
will be established as precisely as possible between individual
scans before establishing an absolute calibration for all data.\\
For channels dominated by the CMB (100GHz to 217GHz), as will be seen 
in next section, the dipole signal 
from the CMB will be used to measure the relative response 
variations on all useful timescales since it is an extended source 
available all over the mission. The amplitude and direction of the 
cosmic dipole has been measured by COBE. It will be remeasured 
independantly by Planck using the dipole from the Earth orbital motion as the 
primary calibrator of the instrument absolute response.\\
For higher frequency channels, mainly dominated by galactic dust 
emission, the calibration will be done by comparison with 
COBE-FIRAS map. The absolute calibration will also be done on the 
whole sky. The relative variations on shorter time scales will be 
measured independantly and are out of the scope of this paper. A short 
summary of the technique is given below. \\
The response variation of a
bolometer depends on a few fixed (or very slowly varying) parameters 
like the thermal conductance or the bias current for exemple. It also 
depends on changing parameters, mainly the temperature of the heat 
sink of the bolometer and the incident power background.  
This last quantity is directly related to the flux from
the sky which is known to first order before the mission or after a first
iteration on the data, and to the temperature of the various stages of the
optics (1.6 K, 4K and the telescope). Since these temperatures are monitored
during the mission, it will be possible to correlate the
response measured as a function of time with the temperature of the
relevant elements and adjust very precisely the parameter of the model 
describing the bolometer chain.  This can be done very accurately using the slow
variations of the different temperatures.  This model with the temperature
measurements (or any other monitored parameter found to affect the
response) can be used to interpolate, if needed, the short time scale
response variations on high frequency channels. Relative variations of the response will
thus be established on many time scales.  A preliminary absolute
calibration will be built after typically a month period and will be
improved until the final full data reduction for the detector is done. 
The bolometer model and the temperature measurements will be used to
interpolate for time scales between one hour and one month.\\
The CMB dipole is a strong known source (using 
COBE data initially or its properties remeasured by Planck at the end 
of the mission) which can be used to monitor the response variations 
over various time scales for all channels where its signal either dominates or is 
clearly detectable. It is thus important to evaluate the 
achievable accuracies for both the final absolute calibration and the 
measurements of the relative response.
\section{Using the CB dipole as a calibration signal for HFI}
The CMB shows anisotropies on all scales tracing the small 
inhomogeneities in the pre-galactic era that lead to the formation 
of structure in the universe. These fluctuations are observed on a surface of last 
scattering as seen from the Earth. They contain high 
spatial frequency terms which are the main objective of the Planck 
mission and low frequency components which are not carrying much 
cosmological information as they are much more variable with the 
observer position (cosmic variance). The dipole term of this cosmological component is 
indistinguishable from the one associated with the peculiar motion of 
the solar system and is therefore a fundamental limit for the accuracy 
to measure this motion.
The CMB cosmologicalanisotropies are thus defined as with no dipole 
component ($\ell=1$ term of the usual decomposition in spherical harmonics of the sky). 
The contribution of higher $\ell$ to the dipole 
measurement is zero when it is done on the whole sky (by 
definition) but is not zero when the dipole is fitted on a fraction of 
the sky.
In this section, we first present the Planck-HFI simulations we used, 
and explore the relative calibration on rings as well as the global 
absolute calibration accuracy.
\subsection{Simulations}
The Planck satellite will orbit around the lagrangian L2 point and 
will therefore follow the Earth orbital motion around the Sun. 
Its scanning strategy consists in rotating the satellite at 1 rpm around its spin
axis which will follow the antisolar direction.  The beam axis, 
located 85$\degr$ away from the spin axis, scans the sky along large circles.  
We define a "circle" as beeing one instantaneous Planck observation of 
1 minute. The
spin axis is relocated about every hour so that each circle is scanned
about 60 times covering the whole sky on six months. These 60 circles are 
averaged into what we call one "ring". The spin axis follows a 
sinusoid trajectory along the ecliptic plane with 6 oscillations per 
year and with full amplitude of 
$\pm 10\degr$ so that the polar caps are not left unobserved.\\
For simplicity and to spare computer time and place, only 84 rings uniformly 
distributed over the sky in a one year mission were 
simulated. Moreover, the removal of slow drifts on circles has to 
be considered together with the process of removing systematic effects 
using redundancies and is therefore out of the scope of this paper. \\
The simulated sky is the sum of 5 contributions:
\begin{enumerate}
    \item CMB cosmological anisotropies, obtained from a standard Cold 
    Dark Matter model: $\Omega_{tot}=1$, $\Omega_{b}=0.05$, 
    $\Omega_{\Lambda}=0$ and $H_{0}=50$km/s/Mpc. The $C_{\ell}$ power 
    spectrum is computed with CMBFAST (Seljak et al. 1996) and the map 
    is generated in a healpix-type all-sky pixelisation with a pixel 
    size of 3.5 arcmin (G\'orski et al. 1998).
    \item The solar system Peculiar Motion dipole (PM dipole) 
    assuming an amplitude of 3.36 mK in the 
    direction $(l,b)=(264.31\degr ,\,48.05\degr )$ (Lineweaver et al. \cite{lineweaver}).
    \item The Earth Orbital Motion dipole (OM dipole)
    assuming an amplitude of about 336 $\mu$K with its 
    maximum at zero ecliptic latitude in the direction of the OM.
    \item Galactic dust, scaled from IRAS 100$\mu$m map using a 
    17.5K blackbody modified by a $\nu^{2}$ emissivity law (Boulanger 
    et al. 1996). 
    \item One realisation of the noise with levels given in 
    table \ref{tab:HFInoise}. These figures were estimated from the 
    sensitivities of the Planck-HFI channels as defined by Lamarre et al. 
    (\cite{lamarre1}). We include a $1/f^2$ contribution to the noise 
    power spectrum with a knee frequency of 10mHz. 
    In order to define the noise level for rings, the instantaneous 
    sensitivity on a circle has been divided 
    by $\sqrt{60}$, assuming that all circles are coadded into a single 
    ring and that the noise is not correlated from one circle to the other. 
    This is not strictly the case given the frequency aliasing produced 
    by the co-addition (Janssen et al. \cite{janssen}, Delabrouille 
    \cite{delabrouille2}). However, it is a rather good 
    approximation (see figure 3 from Giard et al. 1999). Slow drifts at frequencies 
    lower than the spin frequency $f_{\mathrm{spin}}=1/60$Hz are assumed to be removed by 
    a destriping algorithm (Delabrouille \cite{delabrouille3}).
\end{enumerate}
%
%
\begin{table*}[ht]
\caption{\label{tab:HFInoise}
Expected sensitivities of Planck-HFI per detection chain. The last two 
lines give the expected sensitivities per channel for the 
full mission.}
{\small
\begin{center}
\begin{tabular}[b]{|c|c|c|c|c|c|c|}
\hline
Frequency (GHz) & 857 & 545 & 353 & 217 & 143 & 100 \\
\hline
Resolution (arcmin) & 5 & 5 & 5 & 5 & 7.1 & 9.2 \\
\hline
Number of detector & 4 & 4 & 4 & 4 & 4 & 4 \\
\hline
NET$_{\mathrm{CMB}}$ ($\mathrm{\mu}$K.Hz$^{-0.5}$) & 182000 & 3995 & 553 & 182 & 123 & 99\\
\hline
Thermo. temperature sensitivity per ring ($\mathrm{\mu}$K) &  
199000 & 4370 & 605 & 200 & 113 & 80 \\
\hline
NEI (MJy.sr$^{-1}$.Hz$^{-0.5}$) & $269\,10^{-3}$ & $232\,10^{-3}$ & $165\,10^{-3}$ &
$88\,10^{-3}$ & $47\,10^{-3}$ & $23\,10^{-3}$\\
\hline
Intensity sensitivity per ring (MJy.sr$^{-1}$) & $294\,10^{-3}$ 
&$253\,10^{-3}$ & $180\,10^{-3}$ & $96\,10^{-3}$ & $43\,10^{-3}$ & $19\,10^{-3}$\\
\hline
\hline
Thermo. temperature sensitivity full mission ($\mathrm{\mu}$K) & 
36500 & 801 & 111 & 37 & 17 & 11\\
\hline
Intensity sensitivity full mission (MJy.sr$^{-1}$) & $54\,10^{-3}$ & 
$46\,10^{-3}$ & $33\,10^{-3}$ & $18\,10^{-3}$ & $6.6\,10^{-3}$ & 
$2.6\,10^{-3}$ \\
\hline
\end{tabular}
\end{center}}
\end{table*}
\subsection{Ring Analysis}
\begin{figure*}
    \vspace{0cm}
    \hbox{\hspace{0cm}\psfig{figure=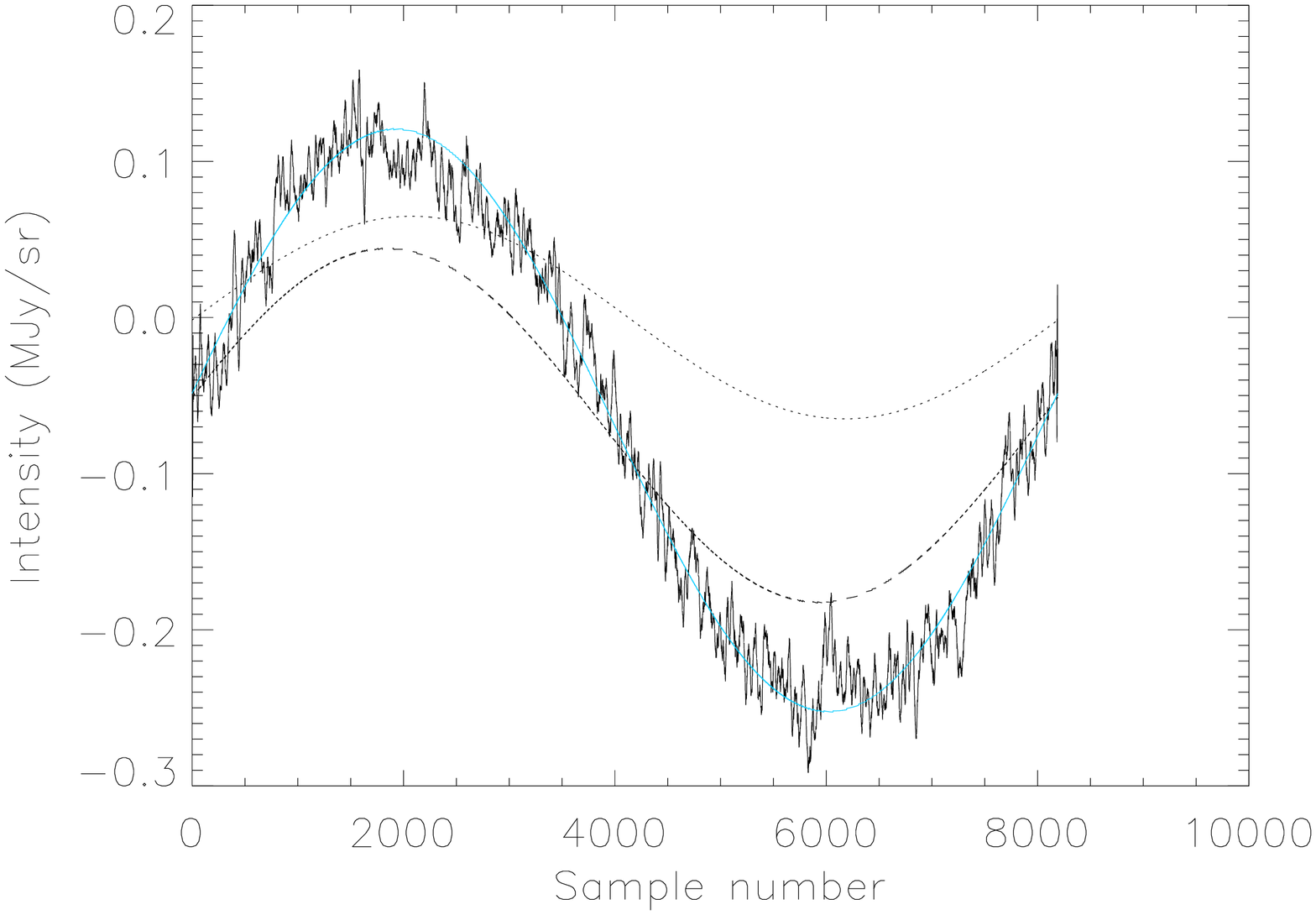,width=8.8cm,angle=90}\hspace{0cm}
    \psfig{figure=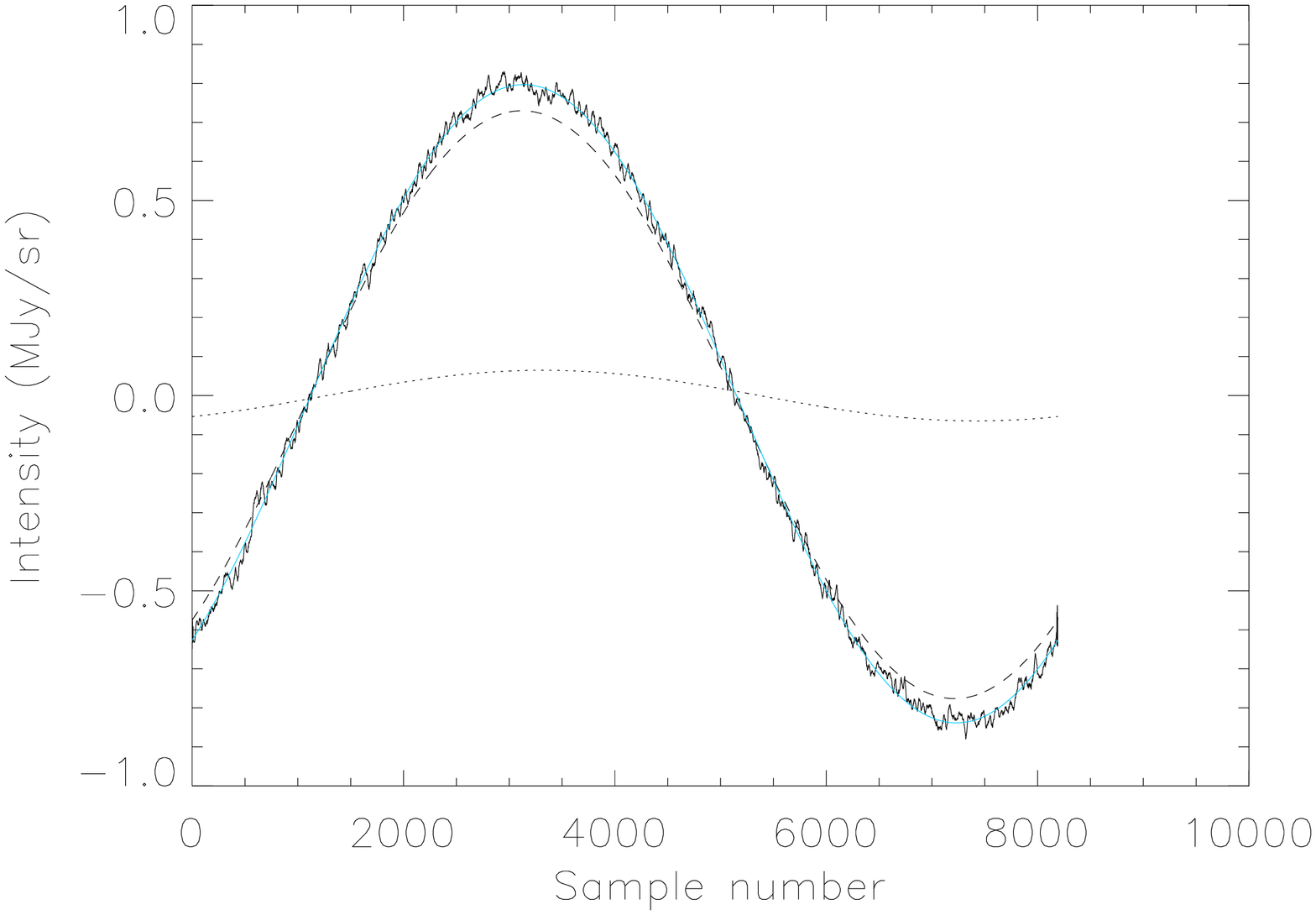,width=8.8cm,angle=90}}
    \vspace{0cm}
    \caption[]{Example of the analysis realised on two rings. The black solid 
    curve is the simulated signal. The OM and PM dipoles are 
    represented by dotted and dashed curves respectively. The grey curve 
    is the OM plus PM motion dipole fitted on the signal.}
    \label{figsimu}
\end{figure*}

For each rings, 
a linear regression is done between the simulated signal 
and the expected dipole signal which is the sum of the PM dipole and 
the OM dipole as shown on figure \ref{figsimu}. The slope of this linear regression gives the 
absolute calibration factor with its uncertainty. The variations of 
this factor gives the relative calibration accuracy. The standard 
deviation of the calibration factor is therefore a good estimator of 
the relative calibration accuracy. The absolute calibration will 
be determined finally on the whole sky map (see section 
\ref{sec:fullcalib} for a first attempt).
A mask on region with IRAS 100$\mu$m emission higher than a cut-off value 
can be applied in order to decrease the contribution from dust 
emission. 
\subsubsection{Relative calibration accuracy}
\begin{figure}
   \hspace{0cm}\psfig{figure=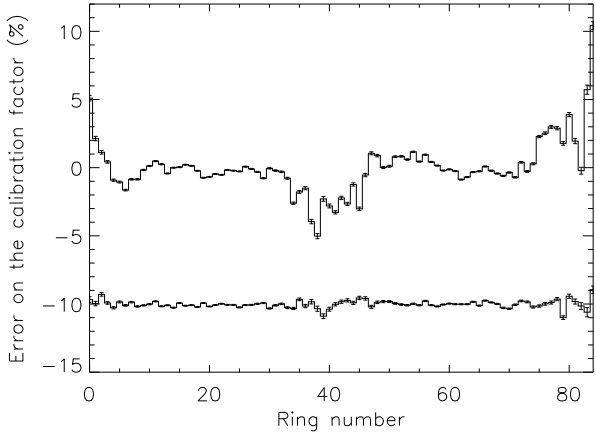,width=8.8cm,angle=90}
   \caption[]{Variations of the calibration factor for a HFI 100GHz 
    channel as a function of the simulated ring number. We have assumed  
    perfectly known OM and PM dipoles characteristics. The lower curve has been 
    computed without CMB cosmological anisotropies and is shifted by 10\% for clarity. 
    Regions where IRAS 
    $100\mu m$ emission greater than 10MJy/sr were removed from 
    the fit. }
    \label{figring100}
\end{figure}

\begin{figure}
   \hspace{0cm}\psfig{figure=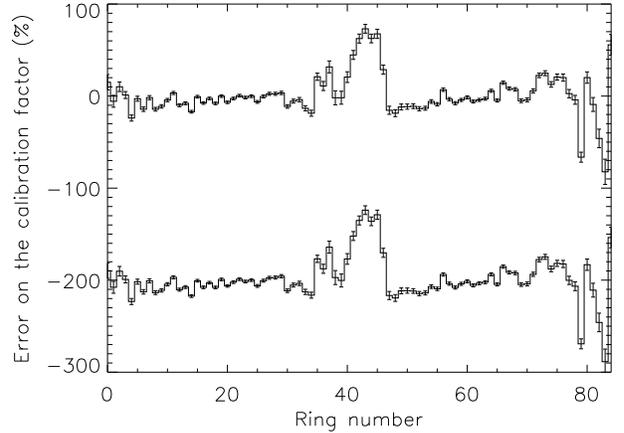,width=8.8cm,angle=90}
    \caption[]{Variations of the calibration factor for a HFI 545GHz 
    channel as a function of the ring number. The legend is similar to 
    figure \ref{figring100}. The lower curve is shifetd by 200\% for clarity. }
    \label{figring545}
\end{figure}

In this analysis, we removed regions with IRAS 100$\mu$m emission 
higher than 10MJy/sr which is quite optimal as will be shown in \ref{dustconta}. 
Figure \ref{figring100} shows the variations of the calibration factor for a 
HFI 100GHz channel along the 
simulated mission assuming known OM and PM dipoles characteristics, with and 
without CMB cosmological anisotropies.
While the PM dipole varies along the mission, the OM dipole 
produces almost the same signal on each rings, as shown on figure 
\ref{figsimu}. The accuracy with which the dipole can be measured 
remains high even when the PM dipole is at its minimum, on 
rings around numbers 40 and 80. When CMB 
cosmological anisotropies are not included, the relative variations of the 
calibration factor are very low, mainly dominated by noise effects. CMB 
anisotropies dominate the uncertainties in measuring the amplitude of 
the dipole on individual ring. The accuracy improves 
with averages on larger number of rings. This part of the error goes 
exactly to zero when the whole sky is used. The 
accuracy of dipole fitting on HFI rings is therefore highly limited by 
the CMB cosmological anisotropies. This fit is equivalent to find the m=1 mode of 
the ring power spectrum $\Gamma_{m}$. The relation from
Delabrouille et al. (\cite{delabrouille1}) links this 
spectrum with the $C_{\ell}$ spectrum:
\begin{equation}
    \Gamma_{m}=\sum_{\ell=|m|}^{\infty}W_{\ell m}C_{\ell}
    \label{eq:delabrouille}
\end{equation}
where $W_{\ell m}$ are window functions related to the beam shape and 
to the ring radius. The dipole term given by $\Gamma_{1}$ 
clearly depends on all multipole $\ell \geq 1$ of the CMB 
cosmological anisotropies. 
This coupling between modes disappear on a full map, indicating that 
the final absolute calibration should be done on the full map of 
Planck data as illustrated in section \ref{sec:fullcalib}. 
Nevertheless, relative calibration between rings at 100GHz is 
expected to be possible at a level of the order of a few percent as 
discussed in \ref{dustconta}.\\
On 545GHz calibration data shown on figure \ref{figring545}, the signal is dominated by galactic 
dust emission and the accuracy of the CMB dipole calibration is about the same 
considering or not CMB cosmological anisotropies. The relative calibration between 
rings at 545GHz is possible at a level of about 15\%.
\subsubsection{Shifted direction of the CMB dipole}
\begin{figure}
    \hspace{0cm}\psfig{figure=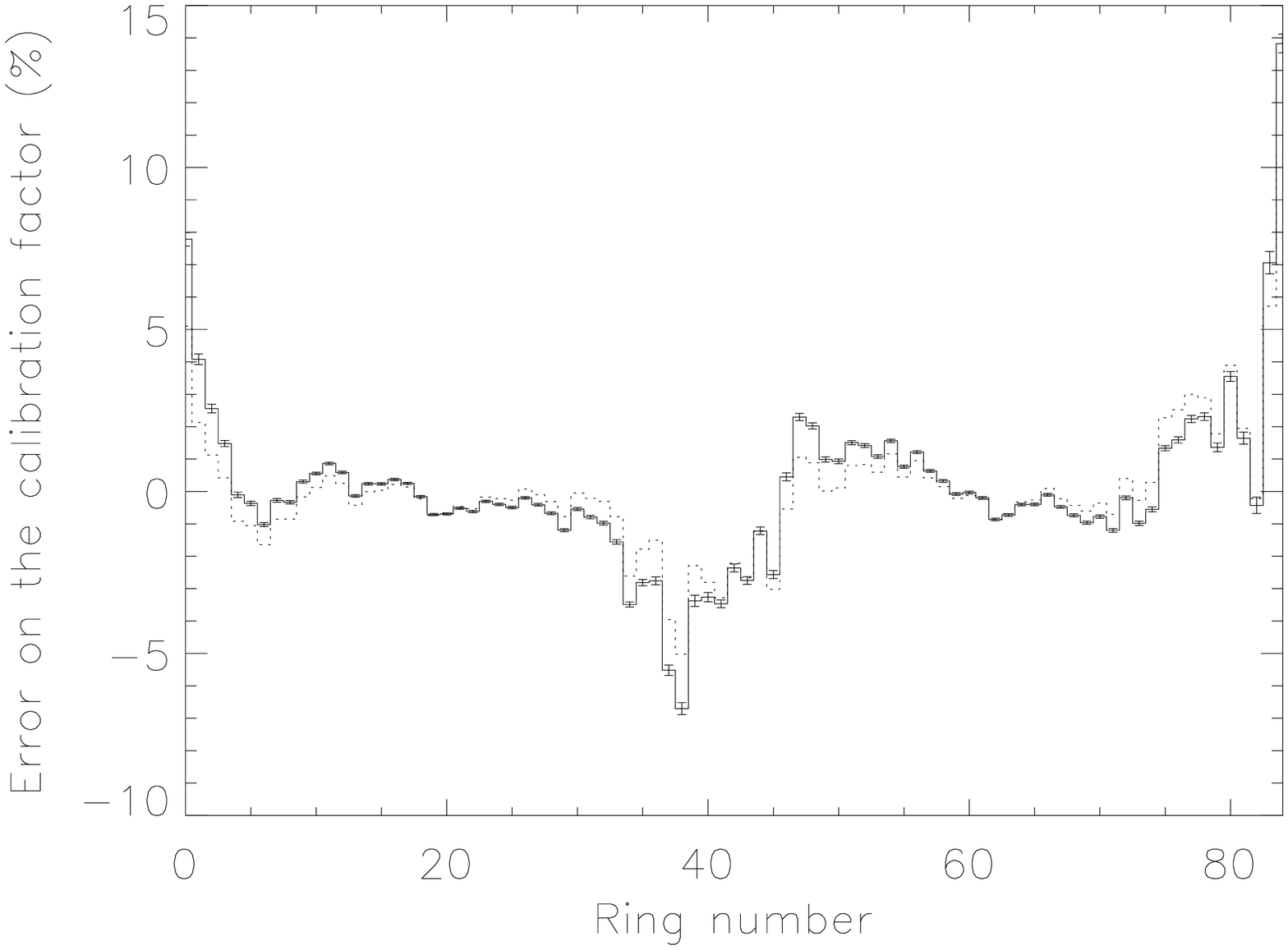,width=8.8cm,angle=90}
    \caption[]{Variations of the calibration factor for a HFI 100GHz 
    channel as a function of the ring number with an error in the 
    PM dipole direction of 30 arcmin. The dotted curve assumed a perfect 
    known PM dipole direction. 
    Regions where IRAS 
    $100\mu m$ emission greater than 10MJy/sr were removed from 
    the fit.}
    \label{figring100shdip}
\end{figure}

The PM velocity vector direction is known with an accuracy of about 14 
arcmin  at $1\sigma$ (Lineweaver et al. \cite{lineweaver}). 
To test the effect of this uncertainty on rings relative calibration, 
we fit the simulated data with a shifted PM dipole by 30 arcmin 
in all directions (about $2\sigma$ from COBE results). 
Figure \ref{figring100shdip} shows the results 
obtain close to the worst case. The accuracy 
on the relative calibration is not highly affected and the calibration factor standard 
deviation is increased by only 15\%.
These results show that a relative 
calibration on rings is still possible in an iterative way:  with a 
first step relative response correction and with enough observation 
time, the dipole direction will be improved 
and therefore the relative calibration will be enhanced.
\subsubsection{Effect of dust contamination}
\label{dustconta}
\begin{figure}
    \hspace{0cm}\psfig{figure=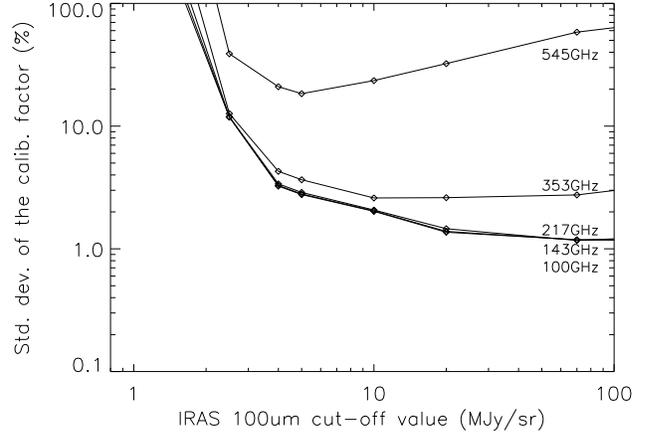,width=8.8cm,angle=90}
    \caption[]{Variations of the standard deviation of the calibration 
    factor (expressed in percentage of the calibration factor) for all HFI 
    channels, except the 857GHz, as a function of the IRAS $100\mu m$ 
    cut-off value. }
    \label{figstd10p}
\end{figure}

The effect of dust contamination on calibration can be tested by 
removing from the dipole fit the data points having an excess emission 
in IRAS 100$\mu m$ map. For very low cut-off 
values, the fit is degraded due to the decrease in the number of data 
points. On the other hand, with a high cut-off value, all the 
emission of dust in our Galaxy is taken into account in the fit and it degrades 
the final accuracy. We therefore expect an optimum value to exist in 
between. 
In order to study this effect, the standard 
deviation of the calibration factor on all rings is 
plotted in figure \ref{figstd10p} as 
a function of the IRAS 100$\mu m$ cut-off.\\
On CMB channels (100GHz, 143GHz and 217GHz), increasing the IRAS 100$\mu m$ 
cut-off value do not decrease the relative calibration accuracy: as 
shown in previous section, this accuracy is limited only by CMB 
cosmological anisotropies. 
A relative calibration accuracy of 1.2\% at 
$1\sigma$ is possible on these channels. On higher frequency channels, 
the emission of galactic dust is more dominant and there is, as 
expected, an optimum cut-off value at a few MJy/sr.
%
%
\subsection{Absolute calibration accuracy on all simulated data}
\label{sec:fullcalib}
\begin{figure}
    \hspace{0cm}\psfig{figure=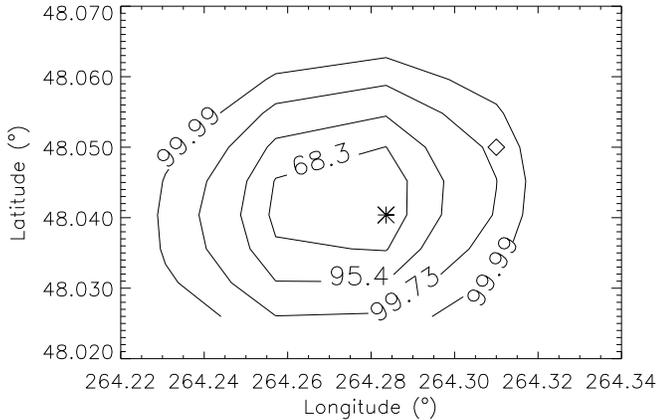,width=8.8cm}
    \caption[]{Contour plot in the $l/b$ plan of the $\chi^{2}$ obtained 
    with the fit of the PM dipole direction, expressed in percentage of confidence level, for a 
    100GHz channel. 
    We assumed known PM dipole amplitude and 
    instrument response for this fit, and no CMB cosmological 
    anisotropies (see text). The asterisk point represents the best fit while 
    the diamond is the input dipole direction. Regions with IRAS 100$\mu m$ 
    emission higher than 15 MJy/sr were removed from the fit.}
    \label{dipole_dir}
\end{figure}

\begin{figure}
    \hspace{0cm}\psfig{figure=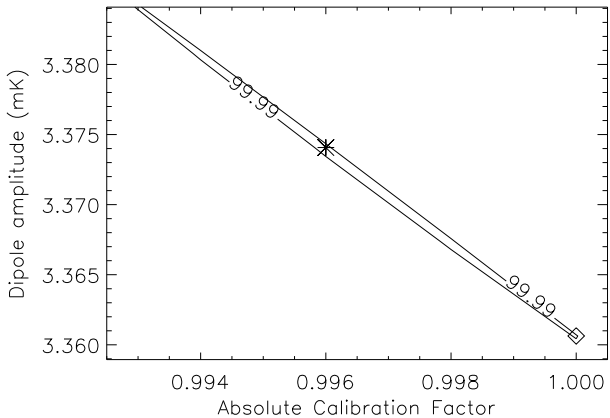,width=8.8cm}
    \caption[]{Contour plot in the $F/A_0$ plan of the $\chi^{2}$ obtained 
    with the fit of the PM dipole amplitude and instrument response, 
    expressed in percentage of confidence level, for a 
    100GHz channel. 
    We assumed a known dipole direction for this fit, 
    and no CMB cosmological anisotropies (see text). 
    The asterisk point represents the best fit while 
    the diamond is the input simulation. Regions with IRAS 100$\mu m$ 
    emission higher than 15 MJy/sr were removed from the fit.}
    \label{resp_amp}
\end{figure}

In order to test the accuracy of absolute calibration on a global map, 
we have fitted simultaneously all simulated rings with a dipole term 
from OM and PM. This is an approximation of a full fit on the whole sky. 
Our procedure nevertheless gives a first 
approach to the global calibration  which will be expensive in computer 
resources. \\
We can write the signal as:
\begin{eqnarray}
    \mathrm{Signal}(l, b) & = &F \times [\mathrm{PM}_{(A_0, l_0, b_0)}(l, b)  + 
    \mathrm{OM(l,b)}\\
    & & + \mathrm{Gal}(l,b) + \mathrm{CMBA}(l,b) ]
\end{eqnarray}
where F is the calibration factor, $A_0$ is the PM dipole amplitude 
($A_0$=3.36 mK),
$(l_0, b_0)$ is the PM dipole direction $(l_0=264.31$\degr$, 
b_0=48.05$\degr$)$, 
Gal is the Galactic signal and CMBA the CMB anisotropy contribution. 
This analysis has been done only on CMB channels (100GHz, 143GHz and 
217GHz).
In order to remove dust contamination effect, we choose an IRAS $100\mu 
m$ cut-off value of 15MJy/sr. Morevover, we remove CMB cosmological anisotropies 
from the fit since, in our case, they introduce a $\ell=$1 term which will 
not be present when using the whole sky. We assume in the procedure that 
we perfectly know only the 
OM dipole so that the calibration is due to this term. 
We first search for the best $F$, $A_0$, $l_0$ and $b_0$ and their accuracy by 
fitting simultaneously the 84 rings signal of one CMB channel. This analysis shows 
that we can separate our problem into two much faster fitting procedures: 
one to get the best calibration factor and PM dipole amplitude,  and the 
other to find the best PM dipole direction and amplitude. This two problems are 
mostly decorelated since a change in the PM dipole direction cannot be 
compensated by a change in its amplitude (or in the calibration 
factor). These two analysis lead to $\chi^2$ contour plot shown in 
figure \ref{dipole_dir} and \ref{resp_amp}. We obtain the following 
results for one 100GHz detection chain:
\begin{itemize}
\item  We can recover the PM dipole direction at better than 1.7 arcmin (Fig. 
\ref{dipole_dir}).
\item  We find $A_0 = (3.374 \pm 0.007)$ mK at 95$\%$ CL and 
\item  F is found to be equal to $0.996 \pm 0.002$ at 95$\%$ CL (Fig. \ref{resp_amp}).
\end{itemize}
This  shows on the one hand that Planck-HFI will be able to make a new 
and accurate determination of the PM dipole direction. 
On the other hand, by combining the error on $F$ and $A_0$ and considering
the OM dipole as an absolute calibrator, we see that we can make an HFI low 
frequency absolute calibration at better than 0.4\%. This first 
attempt is moreover quite pessimistic since it takes into account only 
part of Planck-HFI one year data.
\section{Detectability of the dipole effect in the Far InfraRed 
Background (FIRB)}
The FIRB is due to the integrated emission of 
distant redshifted galaxies. Its SED between 100$\mu$m and 1mm is 
well represented by the following expression (Lagache et al. 1999):
\begin{equation}
    I(\nu)=8.8\;10^{-5}\left( \frac{\nu}{\nu_{0}} \right) ^{1.4} 
    P_{\nu}(T_{0})
    \label{eq:FIRBSED}
\end{equation}
where $\nu_{0}=100cm^{-1}$, $T_{0}=13.6K$ and $P_{\nu}(T)$ is the 
Planck function. This expression has 
been obtained by averaging over multiple clean region, 
almost homegenously distributed on the sky, where dust contamination is 
expected to be negligible. We therefore assumed that it represents the 
SED of the FIRB at rest. The FIRB dipole spectrum can be deduced from its 
monopole spectrum by using equation \ref{eq:2} and \ref{eq:FIRBSED}:
\begin{equation}
    I_{\mathrm{dip\,FIRB}}(\nu)=\beta I(\nu) \left(\frac{x 
    \exp x}{\exp x-1}-1.4 \right)
    \label{eq:FIRBdipSED}
\end{equation}
where $x=h\nu/(kT_{0})$.

\subsection{Signal to Noise ratio}
\begin{figure}
    \hspace{0cm}\psfig{figure=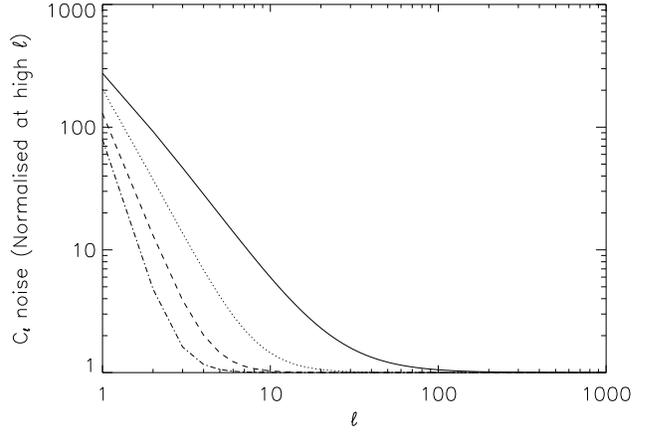,width=8.8cm,angle=90}
    \caption[]{Effect of $1/f$ noise on the noise angular power 
    spectrum $C_{\ell\,\mathrm{noise}}$ normalised at high $\ell$, for 
    a knee frequency of 10mHz. 
    The solid curve correspond to a power $\alpha=1$ (see equation 
    \ref{eq:nep}) while the dotted, dashed and dot-dashed lines are 
    for $\alpha=2,\;3$ and $4$ respectively.}
    \label{cloofnoise}
\end{figure}

In order to evaluate the expected sensitivity of HFI on the FIRB dipole 
signal, we first consider the case of instrumental noise 
having a pure white spectrum. 
As for the CMB dipole, the FIRB dipole will be best detected on a full 
sky map where no aliasing from CMB anisotropies will occur. An 
estimation of the Signal-to-Noise Ratio (SNR) can be done on the $\ell=1$ 
multipole of the angular power spectrum decomposition. 
The expected $C_{1}$ for a dipole signal having an amplitude $A$ is given by:
\begin{equation}
    C_{1}=\frac{4\pi}{9}A^{2}
    \label{eq:dipamp}
\end{equation}
Knox (1997) and Tegmark (1997) have shown that the effect of uniform instrumental 
noise can be accurately modeled as 
an additional random field on the sky, with an angular power spectrum 
given by:
\begin{equation}
    C_{\ell \,\mathrm{noise}}=\Omega_{b} \sigma^2
    \label{eq:clnoise}
\end{equation}
where $\Omega_{b}$ is the  beam solid angle and $\sigma$ is the 
r.m.s. noise per pixel. 
The SNR on $C_{1}$ is therefore equal to $SNR=C_{1}/C_{1 \, noise}$ 
and the results are summarised in table \ref{dethfifdcs}.  
\begin{table*}[ht]
\caption{\label{dethfifdcs}
FIRB dipole level in the Planck-HFI channels and SNR for three 
pixelisations assuming a white detector noise only. We also assumed a 1.17 year mission, square pixels with 
size given by FWHM and bolometer noise given on table \ref{tab:HFInoise}.}
{\small
\begin{center}
\begin{tabular}[b]{|c|c|c|c|}
\hline
Frequency (GHz) & 857 & 545 & 353 \\
\hline
FIRB dipole intensity amplitude (MJy/sr) & $1.6\,10^{-3}$ & $3.4\,10^{-4}$ & 
$4.9\,10^{-5}$ \\
\hline
SNR per pixel & 0.04 & 0.01 & 0.00 \\
\hline
SNR per pixel of $1\degr$ & 0.50 & 0.12 & 0.2 \\
\hline
SNR per pixel of $10\degr$ & 5.0 & 1.2 & 0.2 \\
\hline
\hline
$C_{1\,\mathrm{FIRB}}$ ($(MJy/sr)^{2}.rad^{2}$) & $3.5\,10^{-6}$ & 
$1.7\,10^{-7}$ & $3.4\,10^{-9}$ \\
\hline
$C_{1\,\mathrm{noise}}$ ($(MJy/sr)^{2}.rad^{2}$) & $6.1\,10^{-9}$ & 
$4.6\,10^{-9}$ & $2.3\,10^{-9}$ \\
\hline
SNR full map on $C_{1}$ & 565 & 36 & 1.5 \\
\hline
\end{tabular}
\end{center}}
\end{table*}
It shows that the FIRB dipole can be detected at 857GHz and 545GHz but 
not at 353GHz. Even an averaging 
over $10\degr$ could allow to detect the FIRB 
dipole effects with a significative SNR on these two high frequency 
channels if the detection noise is white.\\
Nevertheless, the HFI will exhibit 1/f noise mainly due to the readout 
electronics (Gaertner et al. 1997, Piat et al. 1997), temperature fluctuations of 
the cryogenic system (Piat et al. 1999) and from far side lobe signal 
(Delabrouille \cite{delabrouille2}). The required knee frequency is 10mHz for HFI in order 
to have enough stability on a timescale of 1 minute corresponding to the spin 
period of the satellite (Piat et al. 1999).
Destriping algorithm can remove an important fraction 
of low frequency drifts. Delabrouille (\cite{delabrouille3}) has shown that 
removing  fluctuations on timescales larger than the spinning period 
of Planck is possible thanks to redundancies. 
A conservative analysis could be made by assuming that all slow components 
of the noise, except frequencies lower than the spin 
frequency, are not filtered. Equation \ref{eq:delabrouille} 
allows to express the ring anisotropy power 
spectrum $\Gamma_{m}$ with the $C_{\ell}$ angular power spectrum on 
the sky. The window functions $W_{\ell m}$ are given by (Delabrouille 
et al. \cite{delabrouille1}):
\begin{equation}
    W_{\ell m}=B_{\ell} \mathcal{P}_{\ell 
    m}^{2}(\Theta)
    \label{eq:wlm}
\end{equation}
where $\Theta$ is the ring angular radius, $B_{\ell}$ is the beam response function (assuming a symmetric 
beam) and $\mathcal{P}_{\ell m}$ are given by the following 
expression where $P_{\ell m}$ are the associated Legendre polynomials:
\begin{eqnarray}
	\mathcal{P}_{\ell m}(\theta) & = &
	\sqrt{\left(\frac{2\ell+1}{4\pi}\right)\frac{(\ell-m)!}{(\ell+m)!}}
	\ P_{\ell m}(\cos\theta), \quad{\rm for\ } m\ge 0, \nonumber\\
	& = & (-1)^{|m|} \mathcal{P}_{\ell|m|}(\theta),\quad{\rm for\ } m< 0, \nonumber
\end{eqnarray}
Equation \ref{eq:delabrouille} can be used to deduce the angular power 
spectrum of the instrumental noise from the Noise Equivalent Power 
(NEP) spectrum. \\
A mode $m$ on a ring has an equivalent frequency bandwidth of $1/(2T_{spin})$ 
where $T_{spin}$ is the spin period. Furthermore, the NEP has to be filtered 
by the beam in order to obtain the noise spectrum projected on the sky. The NEP 
is therefore related to the ring power spectrum of the noise by:
\begin{equation}
    NEP^{2}(f)B^{2}(f)=2T_{\mathrm{spin}}\Gamma_{m}
    \label{eq:nepnoise}
\end{equation}
where $f=m/T_{\mathrm{spin}}$ is the frequency. $B(f)$ is the beam 
filter given by the following expression in the case of a gaussian beam 
shape of Full Width Half Maximum $FWHM=\sigma_b\sqrt{8\ln 2}$:
\begin{equation}
    B(f)=\exp \left[-\frac{1}{2}\left(f \frac{\sigma_b 
    T_{\mathrm{spin}}}{\sin \Theta} \right)^2 \right]
   \label{eq:beam}
\end{equation}
Equation \ref{eq:delabrouille} can therefore be expressed in 
terms of NEP:
\begin{equation}
    NEP^2(f=m/T_{\mathrm{spin}})=\frac{2T_{\mathrm{spin}}}{B^{2}(f)}
    \sum_{\ell=|m|}^{\ell_{\mathrm{max}}}C_{\ell\,\mathrm{noise}}W_{\ell m}
    \label{eq:piat}
\end{equation}
where the sum is done on all accessible $\ell$. The value of 
$\ell_{\mathrm{max}}$ is given by the highest frequency 
$f_{\mathrm{max}}$ transmitted by the 
readout electronics, about 100Hz for HFI that leads to $\ell_{\mathrm{max}} = 
T_{\mathrm{spin}}f_{\mathrm{max}}/ \sin \Theta = 6000$ for a 
ring angular radius $\Theta = 90\degr$.\\
Equation \ref{eq:piat} can be viewed as a matrix multiplication:
\begin{equation}
    NEP^2 = M \times C_{\mathrm{noise}}
    \label{eq:matrice}
\end{equation}
where $NEP$ and $C_{\mathrm{noise}}$ are vectors containing 
$NEP(f=m/T_{\mathrm{spin}})$ and $C_{\ell\,\mathrm{noise}}$ respectively. 
M is a triangular matrix since $\Gamma_{m}$  
depends only on $C_{\ell}$  with $\ell \ge |m|$, and the solution is 
therefore quite easy to obtain. With this method, we get 
$C_{\ell\,\mathrm{noise}}$ assuming a noise spectrum of Planck-HFI given by:
\begin{equation}
    NEP^{2}(f)=NEP^2_0 \left[ 1+ \left( \frac{f_{\mathrm{knee}}}{f} 
    \right)^{\alpha } \right]
    \label{eq:nep}
\end{equation}
where $NEP_0$ is the NEP value in the white part of the noise, 
$f_{\mathrm{knee}}$ is the knee frequency and $\alpha$ a constant. 
The effect of $1/f$ noise on the angular power spectrum is shown on 
figure \ref{cloofnoise} for a knee frequency of 10mHz in 4 different cases 
of noise behaviour. 
The SNR on $C_{1}$ will be degraded by a maximum factor of about 
$270$ which nevertheless allows a detection at more than $2\sigma$ at 
857GHz if foregrounds would be negligible.. 
\subsection{Estimation of the total dipole spectrum}
\begin{figure}
    \hspace{0.5cm}\psfig{figure=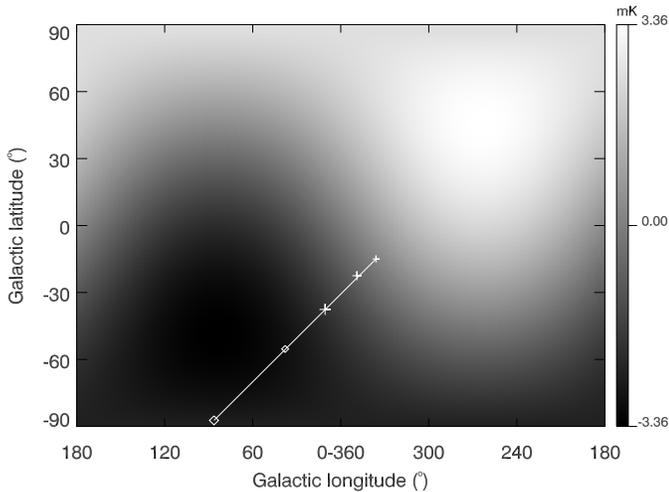,width=7.8cm}
    \vspace{0.6cm}\caption[]{Dust dipole directions superposed to the CMB dipole. 
    The crosses are points obtain with COBE-DIRBE data 
    at 100$\mu$m, 140$\mu$m and 240$\mu$m with increasing size of the 
    symbol, assuming a cut in galactic latitude of $\pm 15\degr$. 
    These points have been extrapolated to HFI 350$\mu$m 
    (857GHz) and 550$\mu$m (545GHz) channels represented by diamonds 
    with increasing size.}
    \label{dustdipdir}
\end{figure}

\begin{figure}
    \hspace{0cm}\psfig{figure=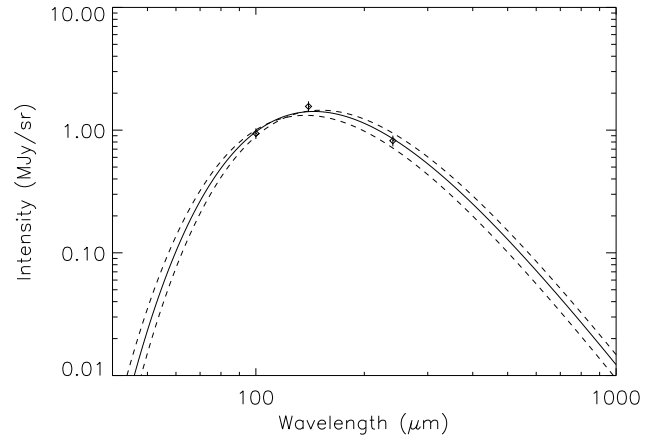,width=8.8cm,angle=90}
    \caption[]{Spectrum of the dust dipole. The diamonds are the points obtain with COBE-DIRBE 
    assuming a cut in galactic latitude of $\pm 15\degr$. The solid 
    line is the best fit with a $\lambda^{-2}$ emissivity leading to 
    a dust temperature of (20$\pm$1)K and a \ion{H}{i} column density of 
    $(5.2\pm1.5)\,10^{19}\mathrm{cm}^{-2}$. Dashed lines gives the 
    upper and lower limits.}
    \label{dustdirspectre}
\end{figure}

\begin{figure}
    \hspace{0cm}\psfig{figure=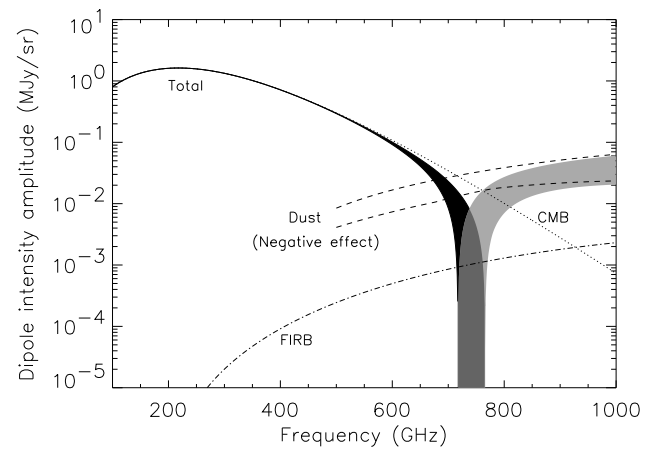,width=8.8cm,angle=90}
    \caption[]{SED of dipoles from the CMB (dotted line), dust 
    (dipole projected on the CMB dipole direction, dashed lines 
    giving lower and upper limits) and the FIRB 
    (dot-dashed line). The solid region gives the spectrum of the 
    total dipole with the uncertainty on the dust dipole, the 
    black one beeing for positive dipole while the grey one is for 
    negative dipole.}
    \label{spectresdip}
\end{figure}

In order to evaluate the effect of calibration and component 
separation, the relative level of FIRB dipole with respect to other 
components has to be evaluated. The total dipole signal on Planck-HFI 
channels is produced by the CMB, the FIRB and also from the 
inhomeneously dust distribution in the Galaxy. \\
The repartition of dust in the Galaxy produces a dipole signal, the 
so-called "dust dipole", that is not due to Doppler effect. This signal has been estimated on 
COBE-DIRBE sky map at 100$\mu$m, 140$\mu$m and 240$\mu$m.
In order to remove strong emission from the galactic plane, a cut in 
galactic latitude or a cut-off in maximum emission level has been 
applied. Both methods leads to about the same results for intermediate 
cut, within less than $20\degr$ on the fitted dipole direction and about 
10\% accuracy on its amplitude. The dipole that dust 
mimic is push toward the galactic south pole as the cut in the 
latitude is increased, or the cut-off in maximum emission is decreased. 
A reasonnable cut in galactic latitude of $\pm 15\degr$ has therefore been applied. 
These data have been used in order to interpolate the dust dipole on 
Planck-HFI 545GHz and 857GHz as shown on figures \ref{dustdipdir} 
and \ref{dustdirspectre}. %
The dipole direction has been 
extrapolated linearly with galactic coordinate, assuming a $\pm20\degr$ 
uncertainty. It seems to go in the direction of the south galactic 
pole, meaning that at higher wavelength only the north/south 
disymmetry will be seen. 
The dust dipole spectrum follows a $(20\pm1)$K blackbody modified by a 
$\nu^2$ emissivity and with a H column density of 
$(5.2\pm1.5)\,10^{19}\mathrm{cm}^{-2}$.
We also assumed that 10\% of this signal can be removed by component separation. 
The resulted dust dipole spectrum with its uncertainty, projected on the CMB dipole axis, is 
shown on figure \ref{spectresdip} together with the CMB, FIRB and 
total dipole.
The FIRB dipole signal is only about 10\% of 
the total dipole dominated by the dust emission one
for frequencies higher than 700GHz, which has two 
consequences:
\begin{enumerate}
    \item To recover the FIRB dipole signal, the dust dipole has 
    to be removed by componenent separation to a 1\% level accuracy 
    (10 times better) which will be difficult to achieve, even 
    on large scales. 
    \item The calibration on the 545GHz and 857GHz channels will be 
    obtained by comparison of the galactic plane emission measured with COBE-FIRAS. 
    The accuracy on these channels will probably be limited by the 
    COBE-FIRAS calibration accuracy which is about 3\%, 
    a factor 3 higher than needed to detect the FIRB dipole effect.
\end{enumerate}
The detection by Planck-HFI of the FIRB dipole is therefore 
unfortunately very difficult, due to component separation and 
calibration accuracies of the high frequency channels.
\section{Conclusion}
We have shown that we can use the PM and OM dipoles observation for relative 
and absolute calibration for Planck-HFI. On CMB channels, we reach a relative 
calibration accuracy of about 1.2\% on rings and an absolute calibration 
accuracy better than 0.4\%  when the whole sky is used. 
Moreover, the CMB dipole direction will be improved by 
Planck. At the end, the Planck-HFI absolute calibration will be 
limited by the uncertainty on the CMB temperature. The next step consist of a 
detail study of systematics 
effects on calibration.\\
The detectability of the FIRB dipole has also been studied and we have 
shown that it will be a difficult challenge due to dust contamination.
%
%
%
%

%
\end{document}